\documentclass [11.5pt]{article}         
\begin{document}

\sloppy

\title{  Permanent Underdetermination from Approximate Empirical Equivalence in  Field Theory: Massless and Massive  Scalar Gravity, Neutrino, Electromagnetic, {Yang-Mills} and Gravitational Theories} %

\author{J. Brian Pitts \\ Department of Philosophy, Department of Physics, and \\
Reilly Center for Science, Technology and Values \\ University of Notre Dame \vspace{.1in} \\Current Address (2016): \\ Faculty of Philosophy, University of Cambridge}

\date{  \emph{The British Journal for the Philosophy of Science}   {\bf 62} (2011), pp. 259-299 \\ doi: 10.1093/bjps/axq014}

\maketitle

\abstract{Classical and quantum field theory provide not only realistic examples of extant notions of empirical equivalence, but also new notions of empirical equivalence, both modal and occurrent. 
A simple but modern gravitational case goes back to the 1890s, but there has been apparently total neglect of the simplest relativistic analog, with the result that an erroneous claim has taken root that Special Relativity could not have accommodated gravity even if there were no bending of light.  The fairly recent  acceptance of nonzero neutrino masses shows that widely neglected possibilities for nonzero particle masses have sometimes been vindicated.  In the electromagnetic case, there is permanent underdetermination at the classical and quantum levels between Maxwell's theory and the one-parameter family of Proca's electromagnetisms with massive photons, which approximate Maxwell's theory in the limit of zero photon mass. While  Yang-Mills theories display  similar approximate equivalence classically, quantization typically breaks this equivalence.  A possible exception,  including unified electroweak theory, might permit a mass term for the photons but not the Yang-Mills vector bosons.  Underdetermination between massive and massless (Einstein) gravity even at the classical level is subject to contemporary controversy.   }

\tableofcontents


\section{Introduction}

 The question whether there exist empirically indistinguishable  but incompatible  theories bears on how tightly empirical constraints from the progress of science might constrain our theorizing.  While the issue of empirical equivalence has been widely discussed, philosophers' discussions often have involved rather thin examples, perhaps generating new theory candidates by de-Ockhamizing (replacing one theoretical entity, perhaps ``force,'' by some  combination of multiple entities such that only that combination plays a role in the theory, such as ``gorce plus morce'' \cite{GlymourEpist}), 
unmotivated deletion of some  regions of space-time or objects therein while the remainder behaves just as in the mother theory, %
and the like \cite{Kukla}.   P. Kyle Stanford argues that resorting to these sorts of examples of underdetermination that philosophers employ, whether algorithmic or not, is a \begin{quote} devil's bargain for defenders of underdetermination, for it succeeds only where it gives up any significant and distinctive general challenge to the truth of our best scientific theories'' by collapsing scientific underdetermination into the familiar and perhaps insoluble problem of  radical skepticism \cite[p. 12]{StanfordUnconceived}. \end{quote}  While there are  interesting issues  in the vicinity of those examples, perhaps more interesting and certainly more novel examples are available in contemporary physics in the context of field theory. To be more specific, scalar gravity,  spinor theories of neutrinos, Maxwell's electromagnetism, Yang-Mills fields, and Einstein's  GR, along with their massive %
relatives, with or without quantization, give examples a number of interesting meta-theoretic phenomena, including the instability of empirical equivalence under change of auxiliary hypotheses \cite{LaudanLeplin,LeplinTotal}. Some of these phenomena  have been anticipated in the philosophical literature but without  such interesting instantiations; others might not have been contemplated previously by  philosophers. %

Empirical equivalence is an issue that plays a key role in arguments about scientific realism.  According to Andr\'{e} Kukla, 
\begin{quote}
[t]he main argument for antirealism is undoubtedly the argument from the underdetermination of theory by all possible data.  Here is one way to represent it: (1)  all theories have indefinitely many empirically equivalent rivals; (2) empirically equivalent hypotheses are equally believable; (3) therefore, belief in any theory must be arbitrary and unfounded. \cite[p. 58]{Kukla} \end{quote} %
On the question whether there exist incompatible but empirically equivalent theories, so that theories are underdetermined by data, a large literature exists, %
but presently there seems to be considerable disagreement on various points. 
By ``empirical equivalence'' I have in mind (unless otherwise qualified) \emph{precise equivalence for all models}---not equivalence that might be broken with further experimental progress or the introduction of differing auxiliary hypotheses, or equivalence that holds in some but not all physically possible worlds.    When I speak of approximate empirical equivalence, once again all models, not just some, are in view.  
During the heyday of logical empiricism, many influential people  %
 denied that distinct and incompatible  but empirically equivalent theories existed \cite{GlymourTheoretical}.   Carnap and Reichenbach had no room empirically equivalent theories, given the verificationist criterion of meaning \cite{CarnapMetaphysics}.  But not only Quine's work \cite{QuineEquivalent},
 but also the revival of scientific realism during the 1960s-70s, led to a revival of belief that distinct and incompatible but empirically equivalent theories exist.  
More recently the view that there do exist rival empirically equivalent theories has been somewhat  widely held \cite{MusgraveRealism,KuklaEquivalence,EarmanCovariance}, in contrast to the earlier positivist view that empirically equivalent theories say the same thing and so are merely linguistic variants. 
This work aspires to address the question of empirically equivalent theories within the context of local relativistic classical and (to some degree) quantum field theories.

Classical and quantum field theories provide philosophically interesting test cases for both approximate and exact empirical equivalence of physical theories. This paper considers approximate empirical equivalence using massless and massive versions of scalar gravity, neutrinos, Maxwell's electromagnetism, Yang-Mills field theories (describing the weak and strong nuclear forces), and Einstein's GR.   By ``massive theories,'' I have in mind the standard particle physics-based sense, with some abuse of language, such that the theories' equations of motion (using the standard potentials, such as appear in the usual Lagrangian density, as variables) have a term that is algebraic and linear in the potential (and no zeroth-order constant term). 
If a  massless scalar field $\phi$ typically satisfies the relativistic wave equation
\begin{equation}
c^{-2} \frac{ \partial^2 \phi}{\partial t^2}  - \frac{ \partial^2 \phi}{\partial x^2} =0, 
\end{equation}
then a  massive scalar field satisfies the Klein-Gordon equation
\begin{equation}
c^{-2} \frac{ \partial^2 \phi}{\partial t^2}  -  \frac{ \partial^2 \phi}{\partial x^2}  + m^2 c^2 \hbar^{-2} \phi=0.
\end{equation}
The coefficient of the algebraic term has been written as $m^2 c^2 \hbar^{-2}$ in anticipation of the use of relativistic quantum field theory.  In units with Planck's constant (reduced by $2\pi$) $\hbar$ and the `speed of light' $c$ set to $1$ (which I use below), it becomes clear why the phrase ``mass term'' is used. For a classical theory, the `mass term' involves not a mass, but an inverse length scale.  In a quantum context, Planck's constant can be used to convert the length-related scale to a mass scale.  The length then corresponds to the Compton wavelength.   The Klein-Gordon equation and other massive field theories with Lagrangian densities have quadratic terms in the potential.  Under quantization (if all goes well), the quanta of fields satisfying the Klein-Gordon equation, such as photons according to Proca's massive electromagnetism, have a nonzero rest mass; then light does not travel at the `speed of light.'  The fact that light need not travel at the speed of light sounds paradoxical, but only if one is misled by the distorting character of older operationalist presentations of relativistic physics.  How fast light actually travels and the  `speed of light' constant $c$ in Lorentz transformations need not be equal.

 At the classical level, at least the simpler  massive theory families have representatives that approximate their massless relatives arbitrarily well empirically  for sufficiently small mass parameters, but the theoretical properties of massive theories can differ greatly from those of their massless counterparts, except in the scalar case (and perhaps the spin $\frac{1}{2}$ case, for which the above discussion would require modification to discuss the first-order Dirac equation).  For example, the massless electromagnetic, Yang-Mills and gravitational theories---the last of these being Einstein's General Relativity (GR) or at any rate a theory with Einstein's field equations---have   mathematically indeterministic field equations and gauge freedom, whereas the massive theories in the standard formulation have deterministic field equations. The fact that the mass of the photon (or vector boson or graviton, \emph{mutatis mutandis}) is a free parameter shows that Proca's massive electromagnetism is not one theory, but an infinite family of theories, one for each choice of value for the photon mass.   This example  therefore exemplifies permanent rather than transient underdetermination.  However,  there is no possibility of identifying supposedly rival theories as really the same theory, because all theories involved (Maxwell's and the various massive electromagnetic theories, for example) are empirically inequivalent. Under quantization, the massive electromagnetic theory remains healthy and  continues to approximate its massless relative arbitrarily well, thus giving an example of underdetermination in a significant quantum field theory.

 The quantized massive Yang-Mills theories (excepting theories with the mass term pertaining to an Abelian sector) and quantized gravitational theories are theoretically problematic, and some  argue that classical massive gravity is also defective.  This phenomenon of a good classical theory that goes bad under quantization resembles or exemplifies Laudan and  Leplin's notion of instability of empirical equivalence under change of auxiliary hypotheses \cite{LaudanLeplin,LeplinTotal}. Here I have in mind taking quantization (or the lack thereof) as something like an auxiliary hypothesis combined somehow with one of the above mentioned field theories, construed in some skeletal form that is neither classical nor quantum.

Explicit and fairly realistic physical theories interrelated through advanced mathematics  should yield insights not as readily available from the traditional discussions of abstract theory formulations   $T_1$ and $T_2$ related by simple logical formulas. 
Earman has made a similar point in the context of determinism:
\begin{quote} If philosophers had spent less time trying to achieve for determinism the superficial `precision' afforded by formal symbolic notation and had spent more time studying the content of physical theories they might have confronted the truly fascinating substantive challenges that determinism must face in classical and relativistic physics. \cite[p. 21]{EarmanPrimer} \end{quote}
The fruitfulness of particle physics for the question of underdetermination reinforces the importance, recently urged by others \cite{BrownPhysicalRelativity,LadymanRoss}, of doing the philosophy of real physics in its detail and richness.


\section{Types of Inexact Empirical Equivalence}

	The most common version of empirical equivalence  discussed by philosophers is the case of exact empirical equivalence for all models of two theories. %
 The potential interest of such a scenario is evident:  obviously there is no chance in any nomologically possible world that experimental progress will resolve the debate, while settling it on theoretical grounds might also be difficult.  However, this scenario runs the risk that the two supposedly rival theories are in fact one and the same theory in different guises.  Such identification was often made by those influenced by logical empiricism.  
A related weaker claim is made today by John Norton, namely, that for theories for which the ``observational equivalence can be demonstrated by arguments brief enough to be included in a journal article \ldots we cannot preclude the possibility that the theories are merely variant formulations of the same theory.'' \cite[p. 17]{NortonUnderdetermine}  Norton evidently has in mind journal articles in the philosophy of science, not physics or some other science \cite[p. 33]{NortonUnderdetermine}. While Norton aims to deny that the underdetermination of theories by data is generic and that philosophers' algorithmic rivals carry much force, I aim to show that there are some serious candidates for underdetermination that arise from within real physics and that have not been discussed much, if at all, by philosophers. Thus there is no real disagreement, except perhaps an emphasis on whether a glass of water is partly empty or partly full.  I make no inductive claim (which Norton would dispute) that these examples imply that all theories are always underdetermined by evidence.  However, the examples available from real physics do seem sufficiently widespread and interesting that it might well frequently be the case that scientific, or rather physical, theories are permanently underdetermined by data.  It is therefore helpful to observe that the physics literature suggests by example several slightly weaker notions of empirical equivalence that, being weaker, are immune to the strategy of being identified as one and the same theory and hence not rivals, yet strong enough that there is no realistic prospect for distinguishing the two theories empirically.

Among philosophers the question has been raised what to make of the many empirically equivalent (or nearly equivalent) theory-candidates or formulations in gravitation, both for Newtonian gravity and for theories employing Einstein's equations  \cite{JonesRealism,LyreEynck,BainNewton}.  Are the several Newtonian (or Einsteinian) theory-candidates just formulations of the same theory, or are they rivals?  If they are rivals, are some of these theories better than others?  Especially because some versions of both the Newtonian and Einsteinian gravities have flat space-time with absolute objects and a gravitational force, while others employ curved space-time, the theory candidates' ontologies and explanatory mechanisms vary rather widely, despite the complete or nearly complete empirical equivalence between the two approaches.  This choice takes up the  issue (discussed by  Lotze, 
Poincar\'{e}, 
and Reichenbach) 
of the equivalence between curved geometry and flat geometry with universal forces.  

The particle physics literature since the 1930s, however, takes up the issue  in a  more detailed and vastly more physically plausible way in the context of gravitation, based on standard 
  principles of special relativistic   field theory such as (of course) Lorentz invariance and the absence of negative-energy degrees of freedom \cite{SliBimGRG,BoulangerEsole} (and see references especially to Fierz and Pauli, Gupta, Kraichnan, Thirring, Halpern, Feynman, Ogievetskii and Polubarinov, Weinberg, Deser, and van Nieuwenhuizen in the former). %

	I point out that this work extends and completes the physical leg of Einstein's double strategy in pursuit of his field equations \cite{EinsteinEntwurfGerman}.  The physical leg of Einstein's double strategy involved, besides an analogy to electromagnetism, a quest for gravitational field equations given by some second-order differential operator equated to the total stress-energy, including gravitational stress-energy, with the gravitational equations by themselves entailing stress-energy conservation. Einstein's equations are in fact logically equivalent to a suitable collection of such laws \cite{Anderson}, corresponding to the conservation of uncountably many energy-momenta \cite{EnergyGravity}.  
 While Einstein retrospectively described his physical strategy as a failure, recent historical scholarship has called attention to its importance \cite{NortonField,Janssen,JanssenRenn}, though without noticing the connection to later particle physics work.  (Particle physicists generally have not noticed, either.)  
 It is worth noting that philosophers' assessments of conventionality in geometry, especially the negative assessment common since the 1970s \cite{PutnamConventionLong},   generally did not keep up with developments in particle physics since 1938. 
It is also noteworthy that  a wholehearted to commitment to thinking of gravitation geometrically tends to constrict the imagination by rendering it nearly impossible to conceive of the massive variants of GR to be considered below.  By contrast, particle physicists' viewing Einstein's equations as describing a self-interacting spin 2 field immediately  suggests that the question whether a massive spin 2 theory might also yield interesting theories of gravity.  This question was indeed pursued from that time   \cite{FierzPauli,TonnelatWaves,deBroglie1,Tonnelat20,TonnelatGravitation,PetiauCR44a,PetiauRadium45}.   %
 Thus the question of massive \emph{vs.} massless spin 2 theories of gravity as a test case for underdetermination has reached philosophers 70 years after it became a common theme in particle physics. A mind-set that obscures interesting questions for seven decades is  worth challenging. It is therefore appropriate to call philosophers' attention to the particle physics tradition. 

In any case, the many different \emph{prima facie} ontologies for gravitation with Einstein's field equations might embarrass scientific realists, who presumably wish to invest the  fields used with (meta)physical significance.  An incomplete list of formulations found in the literature on GR (construed in the physicists' broad and vague sense) reveals a host formulations in terms of different variables and even different numbers of variables.  For curved space-time Einsteinian formulations, there are still many different choices of primitive fields from which to choose, with different numbers of components.
 To mention just some (excluding spinor formulations, for example), one has the typical metric formulation (itself non-unique in Lagrangian density between, \emph{e.g.} the Hilbert $R$ Lagrangian density with second derivatives of the metric and the Einstein $\Gamma\Gamma$ Lagrangian density merely quadratic in the Christoffel symbols,\footnote{In the last few decades \cite{ReggeTeitelboim}, physicists have become less cavalier about discarding boundary terms and regarding Lagrangians differing by divergences as equivalent.  Boundary terms are related to the functional differentiability of the Hamiltonian, so a choice of boundary terms imposes boundary conditions and hence limits the models included by the theory.} and other choices \cite{PonsTraceK} for example, as well as in the choice of variables between the metric $g_{\mu\nu}$, its inverse $g^{\mu\nu}$, and uncountably many densitized relatives  of each); %
 the ADM $3+1$ split with a spatial metric $h_{ij}=g_{ij}$ (lower case Latin indices running from 1 to 3), lapse function $N$ and shift vector $\beta^{i}$  \cite{MTW};  Ashtekar's ``new variables'' for the modern canonical quantum gravity project (a densitized triad-connection version of an ADM split) \cite{JacobsonRomano}; %
 Christian M{\o}ller's orthonormal tetrad formalism \cite{MollerRadiation}; the Einstein-`Palatini'    metric-connection formalism (which is not due to Palatini \cite{FerrarisPalatini}); 
 a metric-connection-Lagrange multiplier formalism that explains why the metric-connection formalism works \cite{Ray}; 
 a tetrad-connection formalism that derives rather than postulates the vanishing of the connection's torsion \cite{PapapetrouStachel}; and the Peres-Katanaev conformal metric density-scalar density formalism \cite{PeresPolynomial}. 
 This last set of variables, though rarely used and little known, turns out to be privileged for Anderson's absolute objects project because it uses  no irrelevant fields (in a fairly well defined sense) and uses only irreducible geometric objects; the failure to use such variables leads to bad performance in inspecting GR for absolute objects \cite{FriedmanJones}.  Roger Jones identified four  formulations of Newtonian gravity; with so many choices of fields available even within geometrical approaches to Einstein's equations, as well as (for example) both geometrical and ``spin 2'' options \cite{LyreEynck,NullCones1} (and the long list above), one sympathizes with Jones's question ``realism about what?'' \cite{JonesRealism}. In what should the scientific realist believe in order to be a realist about gravitation in light of current physics?  Ernan McMullin's claim \cite{McMullinSelective} (with which Stanford sympathizes \cite[p. 16]{StanfordUnconceived}) that mechanics and theoretical physics generally are anomalously difficult for scientific realism  has some basis.

Apart from some exceptions of perhaps little physical importance (such as solutions of an Ashtekar formulation with a degenerate metric, for example), the various sets of variables for GR (broadly construed in the fashion of physicists) 
are empirically equivalent in the sense that all or most solutions of one set of equations are suitably related (not always one-to-one) with solutions in other sets of  variables.  Physicists are generally not tempted to regard the resulting theory formulations  as distinct theories, partly  because their criteria for physical reality are attuned to this mathematical interrelation.  Each description comes with an adequate recipe for distinguishing the physically meaningful from the descriptive fluff, and no further ontological questions are typically asked or answered. Physicists are also quite comfortable with a certain amount of vagueness or merely implicit specificity.  For example, is a given energy condition, such as the weak energy condition  \cite{Wald}, part of GR or not? The answer to that question depends, at least, on whether `realistic' matter fields satisfy the condition; but whether a certain kind of matter is realistic is malleable in light of both empirical factors (such as the apparent observation of dark energy in the late 1990s) 
and theoretical factors (such as recognition that seemingly tame matter fields or quantum fields violate an energy condition hitherto regarded as important) \cite{VisserTwilight}.   GR for physicists is in effect a cluster of theories sharing a hard core including Einstein's equations, while partially overlapping in including or failing to include  various additional claims with various degrees of importance, not unlike a Lakatosian research program \cite{LakatosFalsification} 
 (see also \cite{EllisHandbook}).\footnote{Another possible view is to identify GR with Einstein's equations, and allow the various further specifications to count as sectors within GR.  Such a view has the consequence, it would seem, of making the ontology of GR indeterminate without specifying which sector one has in mind, which seems unhelpful.  It also suggests that if some of the sectors are physically possible, then all of them are; but one might well think that only globally hyperbolic space-times, for example, are possible, or that they are much closer to the actual world than non-globally hyperbolic space-times.  A standard job of a physical theory is to identify physically possible 
and impossible worlds.  Bundling together what might be physically impossible with the physically possible in a single theory thus arguably misses an important role for theories.}   
 Perhaps Arthur Fine would commend to philosophers the physicists'  approach, which sounds something like his Natural Ontological Attitude  that there is no distinctively philosophical question about the real existence of entities employed in scientific theories, so neither realism nor anti-realism is an appropriate doctrine \cite{FineShaky}.  Physicists typically assume  some sort of mathematical equivalence as necessary and sufficient for two formulations to be the same theory (though strict equivalence is not always required). Lawrence Sklar discusses a strategy along these lines, which seems not unreasonable if we have no familiarity with the theory's entities apart from  theory itself \cite{SklarNoumena}.  

If Fine's call to abstain from metaphysical questions goes unheeded, then the variety in choices of  fundamental variables  suggests a variety of  mutually incompatible ontologies and  explanatory mechanisms.  Does space-time really carry a metric only? Does it have a set of four vector fields in terms of which the metric can be defined (and thus make those vectors ``orthonormal'' at the end of the day)?  Does it have a set of orthonormal vector fields with the extra local Lorentz group quotiented out?  (What would it be for such a thing to exist, anyway?)
Does space-time carry  a metric and an \emph{a priori} independent connection that happens to match ``on-shell''(that is, using some or all of the Euler-Lagrange field equations) the torsion-free Levi-Civita connection determined by the metric? Or is the connection simply defined in terms of the metric, so that the modal force of its metric-compatibility and lack of torsion is logical necessity? Similar questions could be asked about electromagnetism, as Julian Schwinger's least action principle, formulated in terms of the  vector potential $A_{\mu}$ and \emph{a priori} independent field strength  $F_{\mu\nu}$   \cite{SchwingerAction}, %
 shows.
Alan Musgrave, it should be noted, does not despair of answering Jones's question regarding what realists should be realists about in gravity \cite{MusgraveRealism}, but a full answer will require more detailed treatment than Musgrave gives.

If one does wish to ask the metaphysician's question about what contemporary physical theories assert to exist, then some criterion for choosing among the many formulations of GR is needed.  On such matters, the Andersonian tradition \cite{Anderson,FriedmanFoundations}, with some friendly  amendments  \cite{FriedmanJones}, is perhaps the best guide available, as suggested above. Anderson insisted on eliminating irrelevant fields.  The friendly amendments insist on eliminating locally irrelevant parts of  fields---using only local sections when global sections are not needed, to use the modern bundle-speak---and using only irreducible geometric objects.    The resulting collection of  fields is such that all geometric objects needed in the theory can be derived from the fundamental fields but not from any smaller set of fields.  The outcome for GR is that the fundamental variables are a conformal metric density (or its inverse) and a scalar density 
 of arbitrary nonzero weight; the metric and connection have no independent existence, but are defined in terms of these two irreducible geometric objects. There is no orthonormal basis of vectors, even with spinor fields present.  

There are various relationships that might obtain between nearly empirically equivalent theories.  The following list is intended to be suggestive rather than exhaustive, but the variety of options that have genuine physical examples is already striking. %
A modal sort of near equivalence is this:  theory $T_1$ has  all the models (or ``worlds'' for variety) of $T_2$, but $T_1$ has some additional models as well.  If expressed in terms of axioms, $T_1$ is logically weaker than $T_2.$  An interesting example would be to consider GR with the possible further requirement of global hyperbolicity \cite{Wald}. Consider GR without the requirement of global hyperbolicity as $T_1$ and GR with global hyperbolicity as $T_2.$  Clearly there is no hope for disproving  $T_1$ on empirical grounds if one is doing science in a $T_2$-world.  One might also consider ordinary quantum mechanics as $T_2$ and Bohmian mechanics, which need not enforce the quantum equilibrium condition, as $T_1$.  A third example takes $T_{2}$ to be Newtonian gravity and $T_{1}$ to be Cartan's variant of it using a space-time with a curved connection \cite{MTW,MalamentNewton,NortonWoes,NortonUnderdetermine}; the difference between these theories strikes me as rather more significant than it seems to Norton.  

Clearly, much depends on how common and interesting the models in $T_1$ but not in $T_2$ are. If most interesting $T_1$-worlds are also $T_2$-worlds, then finding oneself in a $T_2$-world will not even probabilistically confirm $T_2$ much over $T_{1}.$  There might, however, be a sense in which $T_1$ would be noticeably disconfirmed for a scientist in a $T_2$ world if $T_2$ worlds are only a small portion of the worlds of $T_1$. Anthropic considerations might also be relevant if embodied scientists could not exist in some worlds: embodied scientists will certainly not discover that  models incompatible with the existence of embodied scientists are realized in nature. 
 
 A  second kind of modal near-equivalence could arise if each theory has some models not in the other theory, along with some shared models.  For example, GR with global hyperbolicity and GR with asymptotic flatness (such as can obtain for localized sources \cite{Wald}) share some models, while each theory has models that the other lacks.  

A third kind of near-equivalence arises if every model in $T_2$ is diffeomorphic to part, but perhaps not all, of a model of $T_1$.  Some of the most obvious examples, such as might posit that the world began 5 minutes ago %
  or even 6000 years ago, %
 or that certain objects exist only intermittently, %
while otherwise agreeing with conventional history, %
 might seem contrived (but see Kukla for a discussion of the problem of scientific disregard for bizarre theories \cite{Kukla}).  %
 In fact examples of the  phenomenon of a theory's having a model that is a proper part of a model of a related theory,  need not be of the ``ad hoc cut-and-paste variety'' \cite[p. 630]{EarmanSingular}, to use Earman's phrase.  One example is the spin $2$ route to Einstein equations (cited above) with the background metric taken seriously \cite{NullCones1}  as $T_2$, while geometrical GR is $T_1.$ For example, for the Reissner-Nordstr\"{o}m solution for charged spherically symmetric masses in the non-extremal ($Q^2 < M^2$) case, it appears to be possible to include the region between the outer and inner horizons, but impossible to include 
the region within the inner horizon, while taking the flat background metric's null cone seriously as a bound for the effective curved metric.
Thus the realistic spin $2$ approach to Einstein's equations has a solution that lacks a piece present in the geometrical approach---albeit a piece that one is not likely to miss.   A second example takes GR in terms of a metric as $T_2$ and GR in terms of the Ashtekar variables including a connection and a densitized spatial triad, which permit a degenerate metric (that is, with vanishing determinant)  \cite{JacobsonRomano}, as $T_1$.  

All of these kinds of empirical near-equivalence have the property that there is no experiment that can be performed in both theories and such that the results disagree, but they give  different lists of physically possible worlds.  While these sorts of examples merit philosophers' attention, I will set them aside to focus on a less recondite  phenomenon.  However, the phenomenon in question, taken from particle physics, requires comparison not of two theories as is customary, but of one theory and a one-parameter family of rivals.

\section{Approximate Empirical Equivalence}

There is a kind of empirical near-equivalence that is considerably weaker in some respects, involving differences in occurrent properties in similar events in similar models of the compared theories, and yet implying permanent rather than merely transient underdetermination.  	It has seemed at least \emph{a priori} unlikely to some noted physicists that satisfactory physical theories would be isolated, rather than obtainable as limiting cases of a one- (or more) parameter family of theories characterized, for example, by various particle masses \cite{Seeliger1895a,Neumann,BassSchroedinger,DeserMass,GrishchukMass}. %
Particle masses are related to the range of the relevant potential, where the range is a distance scale $\frac{1}{m}$ over which a potential of the form $\frac{1}{r} e^{-mr}$  has the exponential decay factor decay by $\frac{1}{e}$ \cite[p. 598]{Jackson}. (Recall that the `speed of light' $c$ and reduced Planck's constant $\hbar$ are set to $1$ whenever needed.) For $m=0$ the interaction is said to have long or infinite range, as both electromagnetism and gravity are typically held to have.  

Perhaps more to the point than the question about whether theories should be so isolated is the question why  our theorizing about the physical world, given our finite empirical knowledge, should single out just one out of a variety of viable theories with differing particle masses? While Neumann and Seeliger, writing in the 1890s, certainly did not have the concept of particle masses in mind---the link between particle mass and range of the interaction would come later  \cite{deBrogliePhilMag,Yukawa,OkunPhoton1}---the underdetermination point and (especially with Neumann) the exponential form that would later be related to particle masses were already available.  Seeliger writes (as translated by Norton) that Newton's law was ``a purely empirical formula and assuming its exactness would be a new hypothesis supported by nothing.'' \cite{Seeliger1895a,NortonWoes}. 

 Empirically, the photon mass is constrained to be rather small by ordinary standards \cite{GoldhaberNieto,Luo,Tu,GoldhaberNieto2009}, less than $10^{-50}$ grams.  This bound is close enough to $0$ for most practical purposes.  However, this mass, not being dimensionless, is not close to $0$ in a mathematical sense.  It is doubtful that there is a  nonarbitrary sense in which one can say truly that the empirical evidence makes it probable that photon mass is zero.  (A Bayesian effort will be entertained below.)
The question of the photon mass (or the range of the gravitational potential) counts as a natural example, rather than cultured or artificial (to use Norton's classification, based on that for pearls \cite{NortonUnderdetermine}). The contest, however, is not best framed in terms of a pair of theories,  because of the infinite possibilities for the range of the gravitational potential or the value of the photon mass.

The idea of exploring whether a massive theory could work in place of a massless one (or \emph{vice versa}), much as Seeliger proposed, is a commonplace in particle physics. The massless \emph{vs.} massive competition is an especially interesting and well motivated example of the fact, noted by Pierre Duhem, that the curve fitting problem applies in physics.  Two consecutive section headings from the rather familiar part II, chapter 5 make the point:  ``A Law of Physics Is, Properly Speaking, neither True nor False but Approximate'' and ``Every Law of Physics Is Provisional and Relative because It Is Approximate'' \cite[pp. 168, 172]{Duhem}. There are many ways that a given body of data can be fit by a theoretical formula, but Duhem expects that generally a choice of one option will be made.  However,  the competition between massive and massless theories is generally a competition in which both competitors are taken seriously by physicists until reason to the contrary is found.

One can therefore define the relevant concept of approximate empirical equivalence employed here to motivate a novel sort of underdetermination.  The key point is that the empirical equivalence is not merely approximate, and hence perhaps temporary; rather, the empirical equivalence is arbitrarily close and hence permanent.  
 Let $\{ (\forall m) T_m \}$ be a collection of theories labeled by a parameter $m,$ where all positive values of $m$ are permitted.  (One can admit a positive upper bound for $m,$ but that change makes no difference.)   Let $T_0$ be another theory of the same phenomena. 
If the empirical predictions of the family $\{ (\forall m) T_m \}$ tend to those of $T_0$ in the limit $m \rightarrow 0,$ then the family $\{ (\forall m) T_m \}$ empirically approximates $T_0$ \emph{arbitrarily closely}.   
Though $T_0$  is empirically distinguishable in principle from any particular element $T_i$ of $\{ (\forall m)  T_m \},$  yielding merely transient underdetermination between any two theories, $T_0$ is not empirically distinguishable from the entire family.  At any stage of empirical inquiry, there are finite uncertainties regarding the empirical phenomena.  If $T_0$ presently fits the data, then so do some members of  $\{ (\forall m) T_m \}$  for nonzero but sufficiently small $m.$  While scientific progress can tighten the bounds on $m$ towards $0,$ human finitude prevents the bounds from being tightened to the point that all nonzero values of $m$ are excluded while $T_0$ is admitted.  Thus for any stage of empirical science, there will be underdetermination between $T_0$ and elements of $\{ (\forall m) T_m \}$ with $m$ close enough to $0,$ if $T_0$ is still viable. The underdetermination between $T_0$ and  part of  $\{ (\forall m) T_m \}$ is in this sense permanent.  One can never exclude empirically all the $T_m$ theories with $m>0.$

This sort of approximate empirical equivalence has a major advantage over exact empirical equivalence, namely, that a foe of underdetermination cannot avoid the rivalry by identifying the rivals as formulations of the same theory---except perhaps by an \emph{extreme} verificationism
going beyond the mature form of logical empiricism.  The theory $T_m $ for some specific nonzero $m$ is clearly a distinct theory from $T_0,$ incompatible with $T_0,$ making different predictions from $T_0.$


\section{Approximate Empirical Equivalence in Scalar Gravity:  The Neglected Rivalry}

Before addressing the question of the photon mass, it is helpful to consider an analogous simpler question regarding the range of gravitation, which arose in the 19th century for Newtonian gravity, should have arisen for relativistic scalar gravity (but did not in a timely way), and arose in the late 1930s for theories related to General Relativity. 
With the development of Special Relativity it became evident that a relativistic theory of gravity was needed.  Clearly the instantaneous action at a distance in Newton's theory did not agree with Special Relativity.  A local field theory would be ideal.  Eventually, with some help from Einstein,  Gunnar Nordstr\"{o}m had a satisfactory scalar theory \cite{RennGenesis3Intro}, at least prior to the observed bending of light.    This theory was fully in accord with Special Relativity, in the sense of being a local field theory with (at least) invariance under the Poincar\'{e} group of translations and Lorentz boosts and rotations---though in fact the group is larger, as will appear shortly.  Einstein and Fokker expressed this theory in a more geometrical form, so that it yielded an effectively curved spacetime that was conformally flat \cite{EinsteinFokkerCut}.  In other words, the light cones reflecting the speed of light in  Special Relativity were not affected, but the gravitational potential deformed the volume element of spacetime.

  Developments in group theory as applied to relativistic quantum mechanics from the 1930s, such as by Wigner \cite{WignerLorentz}, 
 classified  fields in terms of representations of the Lorentz group with various masses and various spins. Relativistic massive scalar fields, if non-interacting, satisfy the Klein-Gordon equation.
 Given particle physicists' taxonomy in terms of mass and spin, it is natural to look for and to fill in the blanks by considering all the possibilities. (Apart from particle physics, it is easy to fail to notice the gap, which is in fact what generally happened.) 
 Thus when one considers a massless scalar theory of gravity, such as Nordstr\"{o}m's, it is natural to consider a massive variant and to ascertain whether the massless limit of the massive theory is smooth.  If it is, then the massive variant serves as a rival to the massless theory.

  Discussions of Nordstr\"{o}m's scalar gravity, a serious competitor to Einstein's program for some years during the middle 1910s, are said to have shown that even scalar gravity showed the inability  of Special  Relativity (SR)  to accommodate gravitation  \cite[p. 179]{MTW} \cite{NortonNordstrom}. %
Nordstr\"{o}m's theory indeed has a merely conformally flat space-time geometry \cite{EinsteinFokkerCut},
which one can write as 
\begin{equation} g_{\mu\nu}=\hat{\eta}_{\mu\nu} \sqrt{-g}^\frac{1}{2}, \end{equation} 
where $\hat{\eta}_{\mu\nu}$ (with determinant $-1$) determines the light cones just as if for a flat metric in SR.  
Nordstr\"{o}m's theory is invariant under the 15-parameter conformal group rather than just the 10-parameter Poincar\'{e} group standard in SR, whereas massive variants of Nordstr\"{o}m's theory are merely Poincar\'{e}-invariant and hence special relativistic in the strict sense.  As it happens, massive variants of Nordstr\"{o}m's theory were never proposed in a timely way, and indeed not at all (to my knowledge) until one version was proposed, unwittingly,  in 1968 \cite{FreundNambu,DeserHalpern} using universal coupling to the trace of the total stress-energy tensor. Elsewhere I have shown that there is, at least, a one-parameter family of such theories \cite{PittsScalar}. While the kinetic term is just that of Nordstr\"{o}m's theory, 
the mass terms can be written in terms of an effective volume element $\sqrt{-g}$, which  contains the gravitational potential, and also (in the mass term only) the undistorted volume element $\sqrt{-\eta}$ of the flat metric.  For any nonzero real $w$ (including $w=1$ and $w=0$ by  l'H\^{o}pital's rule), a universally coupled massive variant of Nordstr\"{o}m's theory is given by 
\begin{eqnarray}
\mathcal{L}_{mass} =  \frac{m^2}{64 \pi G} \left[ \frac{  \sqrt{-g} }{w-1}   +   \frac{ \sqrt{-g}^w \sqrt{-\eta}^{1-w} }{w(1-w)}  -  \frac{ \sqrt{-\eta}  }{w} \right].
\end{eqnarray}
One can express this mass term as a quadratic term in the potential (naturally) and, typically, a series of higher powers using the expansion $\sqrt{-g}^w = \sqrt{-\eta}^w + 8 w \sqrt{\pi G} \tilde{\gamma},$ where $\tilde{\gamma}$ is the gravitational potential; note that $\tilde{\gamma}$ means something different for each value of $w.$ 
(The case $w=0$ can be treated by taking the $\frac{1}{w}$th root of this field redefinition; the limit gives an exponential function akin to that used in Kraichnan's work on scalar gravity \cite{Kraichnan}.)
The result is
\begin{eqnarray}
\mathcal{L}_{mass} = 
 -m^2   \left[ \frac{ \tilde{\gamma}^2 }{ 2 \sqrt{-\eta}^{ 2w-1} }   +  \frac{ (1-2w) 4 \sqrt{\pi G} \tilde{\gamma}^3 }{3 \sqrt{-\eta}^{3w-1}  }  + \ldots \right].
\end{eqnarray}
This one-parameter family of theories closely resembles the 2-parameter  Ogievetsky-Polubarinov family of massive tensor theories \cite{OP,OPMassive2}, which can also be derived in a similar fashion \cite{MassiveGravity1}. The case $w= \frac{1}{2},$ which conveniently terminates at quadratic order, is the Freund-Nambu theory \cite{FreundNambu}. 
To facilitate comparing apples to apples, rather than apples to oranges, one can use the $w=0$ theory's exponential field redefinition $ \sqrt{-g} = exp(8 \sqrt{\pi G} \gamma) \sqrt{-\eta}$
for every value of $w$, obtaining 
\begin{eqnarray} \mathcal{L}_{mass} = \frac{ m^2 \sqrt{-\eta} }{64 \pi G} \frac{ [ w e^{8 \gamma \sqrt{\pi G}} - e^{8w \gamma \sqrt{\pi G}} + 1-w ]}{w(w-1)} \nonumber \\ =  -\frac{ m^2 \sqrt{-\eta} }{64 \pi G}   \sum_{j=2}^{\infty}  \frac{ (8 \gamma \sqrt{\pi G})^j}{j!} \sum_{i=0}^{j-2} w^i. \end{eqnarray}

The concepts needed for massive scalar gravity were already available in the 1910s.  That is due especially to Neumann and Seeliger's modification of Newtonian gravity in the 1890s with an exponentially decaying potential \cite{Neumann,Seeliger1896,Pauli,North,NortonWoes}. 
  To find a massive relativistic scalar gravity theory, one only needed to do to Nordstr\"{o}m's theory what they had done to Newton's. 
The mathematics was available in the 1910s or 1920s as well   \cite{CottonCR2,Finzi,Fubini1905,WeylReineInfinitesimal,Finzi1,SchoutenConformal,StruikDG,SchoutenRK,LeviCivita}:  one needs two metrics, one of which is flat, the other being conformally related to it. It is simply an accident of history that relativistic massive Nordstr\"{o}m scalar gravity was not proposed in the 1910s, or at least the 1920s when the Klein-Gordon equation appeared. 

 Had massive scalar gravity been proposed at that time, the history of 20th century space-time theory would have been  different in noteworthy ways, because moves that seemed plausible or inevitable would have been recognized as merely optional or plausible, respectively. 
The infinite range/massless cases of Newton and Nordstr\"{o}m are geometrizable.  The equations for Newton's theory can be recast (perhaps with change of content) in terms of a curious but sophisticated geometrical form with the gravitational field absorbed into the connection \cite{MTW,EarmanFriedman}.  The geometrizability of Nordstr\"{o}m's theory  in terms of conformally flat Riemannian geometries was noted above  \cite{EinsteinFokkerCut}. In both cases one learns something important about arguably surplus structure in the theory.  One should not immediately conclude that one has also learned something about \emph{the world}, about gravitation itself, however.  

In  contrast with the massless (infinite range) cases,  massive (finite range) scalar gravities are not geometrizable. For Neumann-Seeliger nonrelativistic gravity, the gravitational potential cannot be fully absorbed into the connection.  The rather large symmetry group of Newtonian gravity \cite{MTW} is thus reduced to the Galilean group.  
The relativistic massive scalar theories involve both the conformally flat metric $g_{\mu\nu}=\hat{\eta}_{\mu\nu} \sqrt{-g}^\frac{1}{2}$  of Nordstr\"{o}m's theory (as geometrized by Einstein and Fokker \cite{EinsteinFokkerCut}) and the flat metric  $\eta_{\mu\nu}=\hat{\eta}_{\mu\nu} \sqrt{-\eta}^\frac{1}{2}$  of SR, as is obvious from the mass terms above (and the suppressed kinetic term).  The symmetry group is thus reduced from the conformal group to the Poincar\'{e} group. 
The two metrics' conformal parts determining the null cones are of course identical. Because there are two metrics present, one has a good argument for the conventionality of geometry, of a sort entertained in advance by Poincar\'{e}  \cite[pp. 88, 89]{PoincareFoundations} \cite{BenMenahemPoincare,FMS,SexlConvention}, as opposed to Eddington's empiricism  \cite{EddingtonSTG,EddingtonNature}:  there just is not any specific fact of the matter about what metrical geometry is yielded by experiments that are sensitive to the gravitational mass term.  (The modal scope of Poincar\'{e}'s argument is broader than just one's favorite theory, such as GR, to which Eddington appealed.  One does not want a philosophy of geometry to impart a spurious necessity to contingent facts about our best current theory  \cite[pp. 848, 849]{Norton}.)  
For the same reasons, strong versions of the equivalence principle are not admissible:  manifestly inertia and gravitation are not the same phenomenon, because inertia is represented by the background structures, whereas gravity is unambiguously represented by the gravitational potential. 

  It is noteworthy, then, how excessive attention to the equivalence principle and geometrization tends to render empirical rivals to Newton's or Einstein's theories inconceivable:  having insisted on a sparse geometrical ontology in formulating these theories, one lacks the resources to construct rival theories. It is then all too easy to regard the theories in question as inevitable, the wave of the future, an assured result of modern progress, or the like.   But it is only the empirical fact of the bending of light by gravity, not any inherent conceptual defect, that made it impossible to treat gravity adequately as a special relativistic theory of a massive scalar field.
 Relativistic gravitation as such does not burst the bounds of Special Relativity by having a larger symmetry group, contrary to claims that have been made  (\cite{MTW,NortonNordstrom}).
 While the mass term (and hence the missing part of the structure of Minkowski space-time) is not demonstrably necessary, it is certainly permitted.   

   Had Nordstr\"{o}m's theory still  been viable by the time that Wigner's classification of Lorentz group representations in terms of mass and spin was widely known, it seems certain that massive scalar gravity would have been considered.  Its neglect until 1968 \cite{FreundNambu}, if not the present (with the partial exception of works by Dehnen and collaborators \cite{DehnenHiggsScalar}, which, however, are focussed only on certain kinds of matter fields, and not fluids, for example), is one of the many disadvantages from the well known \cite{Rovelli} gulf between general relativists and particle physicists. (Helping to overcome this neglect is one reason for attending to the particle physics view of Einstein's equations above.) Massive scalar gravity is an unusual but interesting instance of an interacting massive scalar theory.  Massive scalar (spin $0$) theories have a smooth massless limit, not only classically, but also under quantization  \cite[p. 246]{WeinbergQFT1}.   The precedent that should have been noticed for massive scalar gravity suggests by analogy that one could consider massive tensor gravity as well.

In the actual contingent history (as opposed to a rationally reconstructed one \cite{LakatosFalsification}), Einstein was unaware of Seeliger's work until  after the final GR field equations were known  \cite[p. 420]{EinsteinSpecial}  \cite[p. 557]{EinsteinForster}  \cite[pp. 142, 146]{EinsteinLunarCut} \cite[p. 189]{EinsteinNotes}.    When he did discuss the idea in 1917 (not yet aware of Seeliger's work) \cite{EinsteinCosmologicalCut}, he drew an  analogy between massive scalar gravity and his cosmological constant term, but a spurious one \cite{Trautman,FMS,Treder,NortonWoes,HarveySchucking}
---an error that many would repeat in future years, spawning repeated corrections. The cosmological constant introduces a zeroth order term, not just a first order term, into the field equations.  (This mistake may have been diagnosed first by Otto Heckmann in 1942 \cite{Heckmann,HarveySchucking}, but with little effect.) 
This false analogy tends to hide from view the possibility of a genuine analog to massive scalar gravity, that is, massive tensor gravity, or massive GR more specifically.  Massive GR will reappear below.

\section{Massive Spinor Fields:  The Neutrino Case}

Massive spinor (spin $\frac{1}{2}$) theories also have a smooth massless limit \cite{WeinbergQFT1,DeserMasslessVector}.  Apart from certain significant details,  it was relatively straightforward to give up the traditional default assumption of vanishing neutrino mass in favor of nonzero masses when doing so helped to resolve other neutrino-related puzzles \cite{NeutrinoMassPramana,NeutrinoMassLNP}.  The fact that neutrinos are now believed to be massive, after having been assumed massless, makes the relevance of empirically permitted mass terms the more evident.


\section{Approximate Empirical Equivalence in Electromagnetism}

\subsection{Proca Massive  \emph{vs.} Maxwell Massless Photons }

It is not immediately obvious that the massless limit is smooth for vector (spin $1$) fields, such as electromagnetism, but ultimately the limit is in fact smooth.  
 This question of the photon mass can be considered at both the classical and quantum levels, giving  philosophically interesting test cases for approximate empirical equivalence.  While the usual Maxwell electromagnetism has a massless photon (if one may follow the common practice of borrowing quantum terminology for classical contexts), it is fairly well known that the massive Proca variants exist and approximate the massless theory arbitrarily well for sufficiently small photon mass in both the classical \cite{Jackson,Sundermeyer} 
and quantum contexts \cite{BelinfanteProca,BassSchroedinger,StueckelbergMasslessLimit,BoulwareYM,Slavnov,Shizuya1,Ruegg,GoldhaberNieto2009}. %
 In thermal contexts, where one might expect the third degree of freedom to be relevant, it decouples in the massless limit, so that it takes forever to reach equilibrium; hence equilibrium thermodynamic quantities based on three field degrees of freedom are physically irrelevant and unobservable \cite{BassSchroedinger,GoldhaberNieto}.   

  Because Maxwell's electromagnetism is empirically distinguishable from any \emph{particular} Proca theory (that is, with some given photon mass), and the various Proca theories with different photon masses are also empirically inequivalent, there is no possibility of trivializing the rivalry by regarding the supposed rivals as merely the same theory in different guises.  However,  for any set of observations with finite precision---which is the only kind that human finitude permits at a given stage of inquiry---there exists a range of sufficiently small photon masses such that the massive electromagnetic theories are empirically indistinguishable from the massless theory.    Furthermore, the difference between the massless theory and the massive theories is quite deep conceptually, because only the massless theory has gauge freedom and thus has field equations that mathematically underdetermine the fields' time evolution (assuming that the potential $A_{\mu}$ is used rather than the field strength $F_{\mu\nu}$), along the lines of the hole argument  in GR.  (In comparison to GR \cite{HoleStory}, the analog of the hole argument for electromagnetism is not a very difficult problem,  because space-time point individuation is not at issue due to the purely internal nature of the gauge transformations (lacking derivatives of the fields) and consequent ease of finding the gauge-invariant observable field strength $F_{\mu\nu}$.)   By the same token, the massive Proca theories merely permit charge conservation (which typically holds as a consequence of the field equations for the charged sources), whereas Maxwell's theory enforces charge conservation and so can be coupled only to conserved sources.   Thus the contest between Maxwell's massless electromagnetism and Proca's massive electromagnetisms provides a paradigm case of approximate empirical equivalence: a contest between (or should one say, among) genuine rivals, which cannot be wholly resolved empirically, and on which matters of considerable interest turn.

The most compact and perspicuous way to begin a technical discussion of a classical field theory is to exhibit its Lagrangian density, a function of some fields and their derivatives, such that the space-time integral of the Lagrangian density $\mathcal{L}$, the ``action'' $S$ of the theory, satisfies the principle of least (or perhaps merely stationary) action.  In simple mechanical cases, the Lagrangian is the kinetic energy less the potential energy.  
  The source-free Maxwell field equations (in manifestly Lorentz-covariant form) follow from a Lagrangian density of the form \begin{equation} \mathcal{L}= -\frac{1}{4}  F_{\mu\nu}  F^{\mu\nu}, \end{equation} where the indices are moved using the Lorentz metric $diag(-1,1,1,1),$ $F_{\mu\nu}=_{def} \partial_{\mu} A_{\nu}- \partial_{\nu} A_{\mu}$ is the electromagnetic field strength,  $\partial$ takes the four-dimensional gradient, and repeated indices are summed from $0$ (time) to $3.$   For Maxwell's theory, the vector potential $A_{\mu}$ admits the gauge transformation $$ A_{\mu} \rightarrow A_{\mu} + \partial_{\mu} \phi$$ for an arbitrary function $\phi;$ this transformation makes no observable difference.  This Lagrangian density is manifestly gauge invariant, because it is built from the gauge-invariant field strength only. For the massive Proca electromagnetisms, the Lagrangian density  is \begin{equation}  \mathcal{L}_p = -\frac{1}{4} F_{\mu\nu}F^{\mu\nu} - \frac{m^2}{2} A_\mu A^\mu. \end{equation}  Evidently the $A^2$ term breaks the gauge symmetry in the massive case.  
The resulting Euler-Lagrange field equations are
\begin{equation}
\frac{\partial \mathcal{L} }{\partial A_{\mu} }  - \partial_{\nu} \frac{ \partial \mathcal{L} }{\partial (\partial_{\nu}A_{\mu}) } =  \partial_{\nu} F^{\nu\mu} - m^2 A^{\mu} =0.
\end{equation} 
Whereas Maxwell's theory has $2$ degrees of freedom at each spatial point (written as $2\infty^3$ degrees of freedom), Proca's theories have $3\infty^3$ degrees of freedom. 
  The extra degree of freedom (at each point), however, is  weakly coupled for small photon masses and so is not readily noticed experimentally.  The treatment of the two theories (or theory types) using the Dirac-Bergmann constrained dynamics formalism is straightforward \cite{Sundermeyer}.  The approximate empirical equivalence between Maxwell's theory and Proca's theories for small enough photon masses is preserved under quantization:  massive quantum electrodynamics (QED)  approximates the standard massless QED arbitrarily well, as noted above. 
It follows that in a world with electromagnetism as the only force, it would be impossible for finite beings
to rule out all of the massive electromagnetic theories empirically, and thus impossible to determine empirically whether gauge freedom was a fundamental feature of the electromagnetic laws.  Here I am forgetting about the Stueckelberg formulation \cite{Ruegg,PittsArtificial},
 which shows  that gauge freedom \emph{per se} is not even distinctive of massless electromagnetism: one can have gauge freedom and a photon mass term, unless one bans certain extra gauge compensation fields. The Stueckelberg mass term takes the form  $ -\frac{m^2}{2} (A^{\mu}  - \partial^{\mu} \psi) (A_{\mu}  - \partial_{\mu} \psi)$. A gauge transformation of $A_{\mu}$ is compensated by changing $\psi$:  $A_{\mu} \rightarrow A_{\mu} + \partial_{\mu} \chi ,$ $\psi \rightarrow \psi + \chi.$ The Stueckelberg formulation raises a new set of questions involving exactly rather than approximately empirically equivalent theories, and so will not be discussed here.

A relevant distinction between massive classical electromagnetism and massive quantum electrodynamics pertains to the tendency of classically fixed parameters to acquire quantum corrections. %
Classically one might take the photon mass to be an arbitrary parameter, handed down from above and not susceptible to explanation, but only to empirical determination.  However, in quantum field theory, a small nonzero bare photon mass might acquire large corrections, whereas a vanishing photon mass is forced to stay vanishing by gauge invariance.  Thus in massive quantum electrodynamics, the smallness of the photon mass seems to call for explanation, but no explanation (other than fine tuning) is available.

\subsection{Bayesian Treatment of the Photon Mass}

One might consider whether Bayesian confirmation theory has the resources to say something useful about the problem at hand.  For example, can one show that the probability that a Proca theory is true  goes to zero as the upper bound on the photon mass goes to zero?\footnote{I thank a referee for asking this question.}  Answering this question requires some discussion of plausible prior probabilities for various values of the photon mass.  The massless photon case is special, special enough that it deserves a finite probability all by itself.   Presumably no nonzero value of the photon mass is special, except for those comparable to the upper bound on the photon mass at a given stage of empirical progress. 
Assigning real-valued probabilities to maximally specific hypotheses about the photon mass, other than the massless case, is problematic \cite{SwinburneConfirmation}; this problem will be addressed here informally by taking probability density as basic. 
 Thus there is a mixture of discrete and continuous values.  One can handle the massless case by including a Dirac delta function term $b \delta(m)$ in the probability density, where $b$ is some positive number less than $2$; one recalls that integrating the right half of $\delta(m)$ gives $\frac{1}{2}$ rather than $1.$  It is somewhat less clear what form the distribution should take for finite values of the photon mass $m$ (apart from being effectively $0$ much past the experimental bounds).  

When somewhat similar problems in particle physics (but without   $b \delta(m)$ at $0$) have been treated by Bayesian means---which treatment seems to be rather rare---it has been proposed that the probability density function does not matter terribly much, as long as one avoids cases that strongly favor values near $0$ \cite[pp. 54-56]{DAgostiniDESY}.  Thus a uniform distribution over some finite interval of mass, a triangular distribution bounded above by a downward-sloped straight line, and a half-Gaussian peaked at $0$ gave comparable results.  A distribution sharply favoring values near zero, on the other hand, was judged to give ``\underline{ridiculous}'' results  \cite[p. 56]{DAgostiniDESY}. (Note that this is not the value $0$ itself, which I give a Dirac $\delta$ spike.)  For the photon mass case, it is visually obvious what happens, at least for distributions not strongly favoring values near $0$ (leaving aside the $b \delta(m)$ term).   New experiments tighten the bounds on the photon mass, chopping off the right end (larger $m$) of the probability distribution and scaling up the remainder.  The $\delta(m)$ term gets scaled up but never chopped.  Thus sufficiently vigorous Bayesian updating will
concentrate arbitrarily much of the probability in the $ \delta(m)$ term representing the massless case, leaving the probability that one or another Proca theory is true to approach $0$ as the photon mass's upper bound goes to $0.$  If, on the other hand, one chooses a prior probability distribution for the photon mass which does favor small nonzero values heavily, then Bayesian updating will be all the less effective in undermining commitment to massive photons.  Insofar as one can justify not favoring small nonzero photon masses in the prior probability---a question perhaps worthy of more attention---it  follows that a series of experiments driving the photon mass bound toward $0$ would likewise drive the probability that a Proca theory is true to $0.$

 Whether this result is very significant in practice is open to question, however.  It provides a diachronic rationality constraint on degrees of belief in the Proca family for a Bayesian agent who lives long enough to see arbitrarily strict bounds placed on the photon mass.  But human finitude and perhaps other factors might well ensure that we cannot, even over generations of scientists, drive the bound below some certain finite value.  Thus the arbitrariness in the prior probability might well fail to wash out. One reason might be the energy-time uncertainty relation in conjunction with the age of the universe, from which some estimate a lower measurability bound of roughly $10^{-66}g$ \cite{Tu}.
It is not clear whether one's degree of belief today that the photon has a nonzero mass  ought to be on the order of $.55$ or $10^{-5},$ though few would opt for $.55$ nowadays.  With the serious possibility that progress in tightening the bounds on the photon mass must cease eventually, qualitatively  the same situation (with still more reluctance to accept values near  $.55,$ assuming that a nonzero mass is not detected) might plausibly  still  exist in 500 years. It appears that Bayesianism's ability to formulate interesting questions here perhaps outstrips human ability to answer them.

\subsection{Possible Inductive Lessons about Underdetermination in Particle Physics}

Whereas the massive scalar and massive spinor cases gave no problems in taking the massless limit, care was needed to achieve the same result for the vector case instantiated by Proca's electromagnetism.  A smooth massless limit does obtain, however.  
Having pondered these cases for spin $0,$ spin $\frac{1}{2},$ and spin $1,$ one  might be tempted by induction to draw some  philosophical morals, especially if one is unmoved with surprise by the smallness of the photon mass in massive quantum electrodynamics.  Three  seemingly plausible morals are:  
\begin{enumerate} 
  \item generically there are rival theories that will remain empirically indistinguishable no matter how far empirical inquiry advances, despite the fact that the rival theories give contradictory answers for the same experiment, because theories with slightly different particle masses (or perhaps other parameters) are available. 
  \item  theories that are nearly empirically equivalent classically remain so under quantization, so empirical equivalence is stable under the change of auxiliary hypotheses from classical to quantum.  
  \item   drawing theoretical conclusions to the effect that gauge freedom (and hence mathematical indeterminism) is present in the physics of the real world  is inadvisable,  because contemporary physical theories offer a choice between Maxwell's electromagnetism, a gauge theory with indeterministic equations for $A_{\mu},$ and Proca's theories with deterministic equations and no gauge freedom. 
  \end{enumerate} Doubtless it is inadvisable in general to invest heavily in metaphysical results that are fragile under small changes in physical theory.  One should pause to feel the force of these lessons and develop appropriate expectations for their fulfillment in more advanced contexts. (This corresponds roughly to learning some of the lessons of particle physics into the early 1960s.)  Then one can be appropriately surprised when the lessons appear to fail for more complicated theories.


\section{No Approximate Equivalence in Yang-Mills Theories?}

\begin{quote}
In the early days of Gauge Theory, it was thought that local gauge-invariance could be an `approximate' symmetry.  Perhaps one could add mass terms for the vector field that violate local symmetry, but make the model look more like the observed situation in particle physics.  We now know, however\ldots. \cite[p. 688]{tHooft}  \end{quote}
It turns out that all three of the above lessons might well be  hasty,\footnote{I thank Ikaros Bigi for illuminating discussions and advice on these matters.} 
because matters are much more complicated for both Yang-Mills fields (used for the theories of the weak and strong nuclear forces) and gravity. In both cases most of the relevant results appeared in the early 1970s, though they have generally escaped discussion among philosophers.  Among other reasons that the three lessons are perhaps too hasty, it turns out that quantization can be dangerous to the health of massive theories. 
The preservation of approximate empirical equivalence between the massive and massless cases of electromagnetism  under quantization is something of an accident due to the theories' simplicity, as appears in consideration of the Yang-Mills and gravitational cases from work in the early 1970s.  

 In particular, quantized massive Yang-Mills theory differs from the massless theory even in the limit of vanishing mass \cite{SlavnovFaddeev,Slavnov}; the $m \rightarrow 0$ limit disagrees with the $m=0$ theory by a finite amount in certain observable predictions.  Moreover, quantized massive Yang-Mills 
theory is either non-unitary or not power-counting renormalizable \cite{Wong,DelbourgoTwisk,Ruegg,tHooft}.  %
The massive Yang-Mills theories envisioned are those with a traditional mass term  
 of the form 
$$- \, \frac{m^2}{2} A^i_\mu A^{i\mu},$$ which breaks the gauge symmetry in an explicit  fashion.  This is  
 not the now-standard Higgs mechanism \cite{WeinbergQFT2,SmeenkHiggs,LyreHiggs} %
 for giving an effective mass and finite range to the  weak nuclear force by spontaneous symmetry breaking, by which an interaction between the Yang-Mills vector bosons and the Higgs scalars, after a field redefinition to measure the scalars with respect to a true energy minimum and another field redefinition (in the form of a rotation in abstract space of the vector potentials by the weak mixing angle) yields an effective mass term for the Yang-Mills bosons.  
Nonunitarity  is disastrous because negative probabilities seem unintelligible. %
The lack of power-counting renormalizability seems  less disastrous to some authors nowadays \cite{WeinbergQFT1} than it once did: one settles for an effective rather than fundamental theory and thus admits  that the theory at hand works only up to some definite energy range, after which further terms would be required.  
Under quantization, the massive Proca electromagnetic theories escape this painful dilemma merely because their mathematical simplicity as  Abelian gauge theories (that is, with a gauge group in which the order of two transformations makes no difference to the result), apart from the mass term that breaks the gauge symmetry,
excludes a troublesome term that appears in the non-Abelian Yang-Mills case \cite{Ruegg}.  Electromagnetism is thus too simple a theory to exhibit the dangers that quantization poses to the health of field theories; its atypical simplicity renders it an insufficiently demanding test-bed for philosophical morals of the sort suggested above.  

 This phenomenon involving Yang-Mills theories exemplifies or resembles Laudan and Leplin's notion of instability of empirical equivalence under change of auxiliary hypotheses \cite{LaudanLeplin,LeplinTotal}.  Yang-Mills and massive Yang-Mills field theories are approximately empirically equivalent classically, but this equivalence appears to be violated  at the quantum level. If there is an  essence of Yang-Mills theories that can be exemplified in either classical or quantum form, then one can take this example as a literal instance rather than mere analogy to Laudan and Leplin's phenomenon, which is cast in terms of logically conjoined theories and hypotheses.


\section{Approximate Equivalence in Electroweak Theory:  Yang-Mills Theory with Essentially Abelian Sector}

The underdetermination between the quantized Maxwell theory and the lower-mass quantized Proca theories  is permanent (at least unless a photon mass is detected, in which case Proca wins).  It does not immediately follow that our best science leaves the photon mass unspecified apart from empirical bounds, however.  Electromagnetism can be unified with an $SU(2)$  Yang-Mills field describing the weak nuclear force into the electroweak theory (see, for example, \cite{WeinbergQFT2}).    The resulting electroweak unification  of course is not simply a logical conjunction of the electromagnetic and weak theories; the theories undergoing unification are modified in the process (\emph{c.f.} \cite{Kukla}, chapter 4).   
Maxwell's theory can participate in this unification; can  Proca  theories participate while preserving renormalizability and unitarity?  
Probably they can  \cite{Cornwall,CornwallPRL,Calogeracos,Ignatiev,DuetschSchroer,Ruegg}. 
 Thus evidently the underdetermination between Maxwell and Proca persists even in electroweak theory, though this unresolved rivalry is not widely noticed.  
 There is some non-uniqueness in the photon mass term, partly due to the  rotation by the weak mixing angle between the original fields in the $SU(2) \times U(1)$ group and the mass eigenstates after spontaneous symmetry breaking.  Thus the physical photon is not simply the field corresponding to the original $U(1)$ group, contrary to naive expectations.  There are also various empirically negligible  but perhaps conceptually important effects that can arise in such theories.  Among these are  charge dequantization---the charges of charged particles are no longer integral multiples of a smallest charge---and perhaps charge non-conservation.  Crucial to the possibility of including a Proca-type mass term (as opposed to merely getting mass by spontaneous symmetry breaking) is the non-semi-simple nature of the gauge group $SU(2) \times U(1)$:  this group has a subgroup $U(1)$  that is Abelian and that commutes with the whole of the larger group. Were the electroweak theory to be embedded in some larger semi-simple group such as $SU(5)$, then no Proca mass term could be included \cite{Calogeracos}. 
The dependence of this outcome and others recently discussed on involved physical calculations shows that these are not examples of theories for which empirical equivalence can be demonstrated by a brief argument in a philosophy of science paper---examples which have been a target of Norton's critique \cite{NortonUnderdetermine}.

\section{Is There Approximate Equivalence for Gravity?  General Relativity and Its Massive Variants}

 If Yang-Mills theories qualify the three supposed lessons mentioned above, so that they fail for mass terms for non-Abelian Yang-Mills theories but perhaps do hold for the Abelian sector of non-semi-simple Yang-Mills theories such as the electroweak theory,  the still greater complication of GR renders the three lessons even more contingent upon detailed physical calculation. After a lull from the mid-70s, there has been since the mid-90s  a large and growing literature on massive gravities \cite{OP,FMS,DeserMass,Vainshtein,Visser,Vainshtein2,GrishchukMass,PetrovMass,MassiveGravity1,Zinoviev}, much of it addressing whether they approach (local) empirical equivalence with Einstein's equations in the massless limit and are theoretically healthy, even at the classical level. 
A widely held view since the early 1970s poses a dilemma  \cite{DeserMass,TyutinMass} asserting  that massive gravities either have $5\infty^3$ degrees of freedom (spin 2) and do not agree with Einstein's equations in the massless limit due to the van Dam-Veltman-Zakharov discontinuity \cite{vDVmass1,vDVmass2,Zakharov}, or they have $6 \infty^3$ degrees of freedom (spin 2 and spin 0) and agree empirically with Einstein's theory in the massless limit (at least classically), but are theoretically unhealthy and physically unstable because the spin 0 field has negative kinetic energy.  

 Whereas the Proca theory is the unique local linear massive variant of Maxwell's electromagnetism, the most famous massive gravity with $6 \infty^3$ degrees of freedom, the Freund-Maheshwari-Schonberg massive gravity \cite{FMS,DeserMass}, %
 is just one member (albeit the best in some respects) of a 2-parameter family of massive theories of gravity \cite{OP}, all of which satisfy universal coupling \cite{MassiveGravity1}. Adding a mass term involves adding a term quadratic in the potential; higher-order (cubic, quartic, \emph{etc.}) self-interaction terms might also be present.  The nonlinearity of the Einstein tensor implies, in contrast to the electromagnetic case, that there is no obviously best choice for defining the gravitational potential. While any such definition requires a background metric $\eta_{\mu\nu}$ in order that the potential vanish when gravity is turned off (typically  flat space-time), thus making massive theories bimetric, one can still choose among $g_{\mu\nu} - \eta_{\mu\nu}$,  $\sqrt{-g} g^{\mu\nu} -\sqrt{-\eta}\eta^{\mu\nu}$ (the best choice for some purposes), $g^{\mu\nu} -\eta^{\mu\nu},$  and so on, as well as various nonlinear choices such as $g_{\mu\alpha} \eta^{\alpha\beta} g_{\beta\nu} - \eta_{\mu\nu}$ and the like \cite{OP,DeserQG}.   In some cases the availability of nonlinear field redefinitions might make some expressions that look like  mass term + interaction term with one definition of the gravitational potential, appear as a pure quadratic mass term with another definition; nonetheless the Einstein tensor remains nonlinear, no matter what definition of the potential is used.  By contrast, the linearity of the Maxwell field strength tensor makes it natural to have a mass term that is also linear in  $A_{\mu}$ in the field equations (and hence quadratic in $A_{\mu}$ in the Lagrangian density).  While one can explore introducing nonlinear algebraic terms in $A_{\mu}$ describing self-interactions in electromagnetism, such terms induce acausal propagation if not chosen carefully \cite{ShamalyCapri}.

 Whether massive gravities are viable even at the classical level remains a matter of debate in the physics literature. 
The majority view is that they are not, but this view has lost the near-consensus status that it once had.  
 Evidently intuitive expectations about the ease of constructing approximately empirically equivalent theories to GR are threatened by devils in the details. %

One possibility worthy of exploration is whether the methods of PT-symmetric quantization can help.   PT-symmetric quantization has  exorcised the vicious ghosts thought to inhabit some theories according to more traditional analyses \cite{Mostafazadeh,Bender,BenderMannheim}, though the resulting theories sometimes have surprising phenomenology.  Might PT-symmetry be helpful for massive gravity or for a \emph{prima facie}  non-unitary  \cite{DelbourgoTwisk}   massive Yang-Mills theory? 

 It is also noteworthy that there are examples from fluid mechanics of theories with negative energy modes that do not yield  instability in an unphysical fashion; such theories have Hamiltonians that are not of the typical separable form of a kinetic energy built from momenta plus a potential energy built from generalized coordinates \cite{Morrison}.  Massive versions of GR have nonseparable Hamiltonians. Thus it seems a bit quick to reject massive gravities simply because of a wrong-sign degree of freedom, without more detailed analysis \cite{MassiveGravity1}. The majority view's reliance on first-order perturbative arguments regarding a theory where nonlinearity in the Hamiltonian constraint should be important, also might give one pause.  The question seems to be not quite settled, therefore.  The role of such theories in assessing underdetermination is a worthwhile motive for pursuit, even if the odds are somewhat long.  Indeed, comparing the literature on massive gravity to works in plasma physics, where negative energy degrees of freedom are routinely encountered, one notices among plasma physicists much less tendency to reject such theories without detailed investigation.  Plasma physicists do not conclude, much less assume, that negative energy degrees of freedom are always automatically disastrous; the demonstrable physical relevance of such theories in plasma physics does not permit rejecting such theories automatically.  Instead plasma physicists  have developed criteria, including looking for resonances and taking into account the magnitudes and signs of the modes in question, that indicate instability \cite{WeilandWilhelmsson}.  

Besides plasma physicists, mathematicians also study theories with positive and negative energy degrees of freedom and fail to find generic instability \cite{Berti}.  Indeed various theorems have been proven (for finitely many degrees of freedom) to the effect that, especially but not only in the absence of resonances,  periodic solutions tend to persist when small perturbations are made about theories with periodic solutions; the absence  of negative energy degrees of freedom is not assumed in some of these theorems.  It seems not to be expected that the situation becomes radically worse in field theories; the question is a subject of current research.  It is therefore doubtful that the issue of massive gravity was settled in the early 1970s.  Of course it might well be the case that massive gravities are unstable for substantially the reasons usually mentioned; if there is perhaps no proof, there is certainly ground for suspicion.

Besides the question of stability, massive gravities, being bimetric, are also susceptible to causality problems if the relationship between the two metrics' null cones is not correct; sometimes it is not, %
at least not 
 without help \cite{MassiveGravity1}. Though matter sees only the effective curved metric and gravity only barely sees the flat background metric due to the smallness of the graviton mass, these theories are only Lorentz-covariant (or covariant under the 15-parameter conformal group in the case of massless spin $0$).  Thus the usual special relativistic arguments about superluminality in one frame implying backwards causation in another frame are applicable. At the quantum level it would be awkward at best to impose  Lorentz-covariant equal-time commutation relations when there is no reason to expect events that are space-like separated (with respect to the background metric) to be causally independent.  
Such causality problems bear some resemblance to those encountered in the 1960s with spin $\frac{3}{2}$ fields 
 \cite{VeloZwanziger2}.  Adding gauge freedom and then judiciously restricting it seems likely to cure the problem of acausality \cite{MassiveGravity1}.  
 Massive variants of GR are of special interest as foils for GR concerning general covariance \cite{PittsArtificial}.

While the above discussions have aimed to survey longstanding issues in particle physics, there are various recent developments also of interest.  In recent years it has been found that if one is willing to introduce a cosmological constant term, in effect using a background space-time of constant curvature, then there are opportunities for a smooth massless limit for massive gravitation \cite{Kogan}. 
Supergravity theories involve the existence of at least one  field of spin $\frac{3}{2}.$  
 There is also a discontinuity in the massless limit for spin $\frac{3}{2}$ vector-spinors \cite{DeserKayStelle}; more recently it was found that a background of constant curvature can make the massless limit smooth \cite{DeserWaldronSuper}.   One might take the view that such alternatives, by leaving the  world of fields in flat space-time, are less simple, and hence perhaps less probable or otherwise less worthy of attention than a case of underdetermination in flat space-time would be. 
On the other hand there are recent surprising indications that highly symmetric ($N=8$) supergravity in four space-time dimensions is ultraviolet-\emph{finite} (better than renormalizable) at least to the fourth loop order \cite{BernFinite4} 
and perhaps to all orders, so spin $\frac{3}{2}$ fields seem more likely to exist in reality than they did previously.  (This sort of work is a reminder that real scientific progress can be made by humans in contexts that would not be a challenge for an ideal Bayesian agent, who is logically and mathematically omniscient \cite{GarberOldEvidence}.)
 At any rate there is a lively current literature on subtle moves that do or might yield a smooth massless limit even for some spins higher than $1,$ but a brief summary would be both difficult and premature. Discontinuity was found some time ago  for higher spin fields as well, such as spin $\frac{5}{2}$ and spin $3$ \cite{Berends5Halves,BerendsSpin3}, in case such fields really exist. String theory, at least, has a rich store of higher-spin fields.    Rather than aim for exhaustive discussion, this paper aims mainly to call attention to longstanding features of particle physics that are relevant to underdetermination, while pointing the way toward a more active philosophical engagement with such literature in the future.


\section{Conclusion}  

The  provisional character of some conclusions involving more complicated field theories  indicates that in these cases, whatever the ultimate outcomes,  the physical details sometimes have important philosophical consequences.  These examples therefore do not provide strong support for claims that underdetermination is generic throughout science.  They do, however, provide  support for the claim that there might well be interesting cases of permanent underdetermination, even in our contemporary best science, such as the  electroweak quantum field theory, as well as in simpler theories.  The simpler the field theory, the easier it is to have underdetermination.  Higher-spin fields are more likely to involve either dangerous wrong-sign degrees of freedom or a discontinuous massless limit.

Consideration of a variety of theories (and formulations thereof) of contemporary physical relevance has yielded a variety of insights pertaining to the  general philosophy of science. The question of empirical equivalence and underdetermination of theories by data proves to have highly nontrivial examples, as well as surprising failures, when one looks  into classical and quantum field theory.  Thus philosophers need not rely  on thin contrived examples.  Particle physics is important for general philosophers of science, as well as space-time theorists; it is not just for philosophers of quantum mechanics. One can add these reasons to Redhead's list of reasons for philosophers to study quantum field theory \cite{RedheadQFT}.  Comparing some contemporary physical theories also suggests apparently novel sorts of underdetermination.    There is a surprising degree of dependence on the physical details, including difficult calculations in quantum field theory, for approximate underdetermination.  Underdetermination can be broken in surprising ways when  auxiliary hypotheses are changed, much  as Laudan and Leplin have anticipated.   The kinds of underdetermination  discussed here are immune to trivialization as just linguistic variants of the same theory.    Underdetermination might not be ubiquitous in contemporary particle physics, but there are enough interesting  examples and types of it that the subject remains of considerable interest.  

There are plausible versions of scientific realism that are not threatened by the examples from particle physics discussed here.  For example, Ernan McMullin writes that 
\begin{quote}
[t]he basic claim made by scientific realism \ldots is that the long-term success of a scientific theory gives reason to believe that something like the entities and structure postulated by the theory actually exists.  There are four important qualifications built into this:  (1)  the theory must be successful over a significant period of time; (2) the explanatory success of the theory gives some reason, though not a conclusive warrant, to believe it; (3) what is believed is that the theoretical structures are \emph{something like}  the structure of the real world; (4) no claim is made for a special, more basic, privileged form of existence for the postulated entities. [footnote suppressed]  These qualifications:  ``significant period,'' ``some reason,'' ``something like,'' sound very vague, of course, and vagueness is a challenge to the philosopher.  Can they not be made more precise?  I am not sure that they can; efforts to strengthen the thesis of scientific realism have \ldots left it open to easy refutation. \cite{McMullinRealism} \end{quote}

On the other hand, one can imagine more ambitious sorts of scientific realism, to the effect that rival theories will always be in principle empirically discriminable, and will be discriminated in the not-too-distant future,
that are threatened.  
A determined scientific realist of such an ambitious type could respond to the cases from particle physics by noting that sometimes evidence has broken the underdetermination, as in the neutrino and Yang-Mills cases and arguably the gravitational case, and that these cases give some comfort to the expectation that either further facts or the instability of empirical equivalence under change of auxiliaries will always ensure that empirically distinguishable predictions be made.  
While such a response is possible, it appears to amount to an expectation of routinely being mildly lucky.  In some cases the underdetermination has indeed been resolved, but it is not at all clear what motivates the determined scientific realist's hope that it will \emph{always} be resolvable, or resolvable on a time scale that makes us reasonably close to learning the truth.  It is difficult to think of a non-Hegelian reason for expecting to be mildly lucky routinely in the required way.  Presumably some sort of scalar gravity could have been true, and gravity would not have bent light as in GR.  The underdetermination case for scalar gravity was not resolved, but merely bypassed; but what right would a scientific realist have to expect such deliverance? 
The underdetermination case for the neutrino was resolved, but  in such a fashion that the widespread belief in massless neutrinos was rejected.  If the scientific realist is comforted by the thought that eventually the underdetermination was broken, there is also a warning that a widely held assumption of masslessness has proved false. If the data speak, they might not do so for a long time, during which interval widespread scientific belief might be mistaken.  
 The electroweak theory might well permit a mass term for the photon.  Does the scientific realist have good reason to expect a further unification using a group like $SU(5)$ that blocks such a mass term?

The no-miracles argument for scientific realism will not help.  It is evident that if a Proca theory for small photon mass (or rather its quantum successor, here and following) is true, it is not a miracle that Maxwell's theory works so well; likewise if Maxwell's theory is true, it is no miracle why Proca theories with small photon masses work so well. The theories are sufficiently alike theoretically as well as empirically; the no-miracles argument has already done its work. But what further reasons does the optimistic scientific realist have?  Routine good luck, though characteristic of the fictional life of Inspector Clouseau, is just the sort of thing that it is not rational to expect in real life by ordinary (non-Hegelian) inductive principles.   One could perhaps fall back upon the principle that one philosopher's \emph{modus ponens} is another's \emph{modus tollens}, or the subjectivity of Bayesian prior probabilities, as licensing rational intransigent commitment to certain strong forms of scientific realism.  But surely one ought to weigh the force of the examples from particle physics against the force of arguments for such  optimistic scientific realism.  As I read particle physics, the weight of the examples of underdetermination above is fairly strong.  In any case one must read particle physics in order to ascertain the force of the examples to be weighed.

\section{Acknowledgments}

I thank Ikaros Bigi, Don Howard, Katherine Brading, and John Norton for helpful discussions and Henriette Elvang for bibliographic assistance.




\end{document}